\documentclass[10pt,amsmath,amssymb, aps, prd,twocolumn,showpacs]{revtex4-1}

\usepackage{graphicx}
\usepackage{dcolumn}
\usepackage{bm}
\usepackage{amssymb}
\usepackage{amsmath}
\usepackage{verbatim}
\usepackage{mathrsfs}
\usepackage{amsfonts}
\usepackage{latexsym}
\usepackage{epsfig}
\usepackage{color}
\usepackage{graphicx,subfigure}
\usepackage{units}
\usepackage{array}

\newcommand{\Msun}{{\rm M}_{\odot}}
\newcommand{\sigmav}{\langle\sigma v\rangle}
\begin{document}

\author{Alma X. Gonzalez-Morales}
\affiliation{Santa Cruz Institute for Particle Physics and Department of Physics, University of California,
Santa Cruz, CA, 95064, USA}

\author{Stefano Profumo}
\affiliation{Santa Cruz Institute for Particle Physics and Department of Physics, University of California,
Santa Cruz, CA, 95064, USA}

\author{Farinaldo S. Queiroz}
\affiliation{Santa Cruz Institute for Particle Physics and Department of Physics, University of California,
Santa Cruz, CA, 95064, USA}

\preprint{APS/123-QED}

\title{Effect of Black Holes in Local Dwarf Spheroidal Galaxies on\\ Gamma-Ray Constraints on Dark Matter Annihilation}

\date{\today}

\begin{abstract}
Recent discoveries of optical signatures of black holes in dwarf galaxies indicates that low-mass galaxies can indeed host intermediate massive black holes. This motivates the assessment of the resulting effect on the host dark matter density profile, and the consequences for the constraints on the plane of the dark matter annihilation cross section versus mass, stemming from the non-observation of gamma rays from local dwarf spheroidals with the Fermi Large Area Telescope. We compute the density profile using three different prescriptions for the black hole mass associated with a given spheroidal galaxy, and taking into account the cutoff to the density from dark matter pair-annihilation. We find that the limits on the dark matter annihilation rate from observations of individual dwarfs are enhanced by factors of a few up to $10^6$, depending on the specific galaxy, on the black hole mass prescription, and on the dark matter particle mass. We estimate limits from combined observations of a sample of 15 dwarfs, for a variety of assumptions on the dwarf black hole mass and on the dark matter density profile prior to adiabatic contraction. We find that if black holes are indeed present in local dwarf spheroidals, then, independent of assumptions, (i) the dark matter interpretation of the Galactic center gamma-ray excess would be conclusively ruled out, (ii) wino dark matter would be excluded up to masses of about 3 TeV, and (iii) vanilla thermal relic WIMPs must be heavier than 100 GeV.

\end{abstract}


\maketitle

\section{Introduction}
One of the biggest open questions of modern cosmology lies with the nature of Dark Matter (DM). DM constitutes approximately $23 \%$ of the energy-density of the Universe, or about 4/5 of the matter density, but its existence has not been confirmed by any experimental means besides gravitational effects. The most compelling particle DM candidates are Weakly Interacting Massive Particles, or WIMPs, which arise in a wide variety of well-motivated theories beyond the  Standard Model, and which can naturally account for the observed dark matter abundance in the framework of standard thermal freeze out (see Ref.~\cite{2010pdmo.book....3B} for an extensive overview on particle dark matter).

A few of the direct detection experiments have observed some excess events consistent with WIMP scattering off of ordinary nuclei, for example COGENT, DAMA, CREST, and CMDS \cite{Aalseth:2012if,Bernabei:2013cfa,Angloher:2011uu,Agnese:2013rvf}. Other experimental collaborations, however, including XENON and LUX, have not confirmed any such excesses, ruling out most of the WIMP parameter space where the signals might arise \cite{Aprile:2012nq,Akerib:2013tjd}. At present, there is no clear evidence for a positive signal from DM direct detection.

The so-called indirect detection of dark matter, i.e. the observation of DM annihilation products, is another promising avenue. Indirect detection has the potential to not only conclusively demonstrate the existence of dark matter, but also to characterize it as a particle in some detail. Indirect detection searches in the Galaxy have found tentative signals in  gamma rays \cite{Hooper:2011ti,Abazajian:2012pn,Hooper:2012sr} and in cosmic ray data \cite{Boezio:2008mp,Ibarra:2013zia} which might be explained by annihilation of WIMPs in the Galactic halo \cite{Cembranos:2012nj}. All of those claimed signals possess, however, known astrophysical counterparts that provide rather compelling, and simpler, explanations \cite{Hooper:2008kg,Profumo:2008ms,Grasso:2009ma, Profumo:2012,Carlson:2014cwa}. In addition, indirect detection searches for a signal from DM annihilation in galaxy clusters and dwarf galaxies have not found any significant excess  \cite{::2013ufa,Storm:2012ty,Ackermann:2011wa, Ackermann:2013yva}.

 The Fermi collaboration has recently reported  $\gamma$-ray observations of 25 Milky Way (MW) dwarf spheroidal satellite galaxies based on 4 years of data \cite{Ackermann:2013yva}. No significant excess has been detected, yielding some of the tightest constraints on the plane defined by the WIMP pair-annihilation cross section versus mass, ($m_\chi, \langle \sigma v\rangle$)  \cite{Ackermann:2013yva}. The results are robust against uncertainties in the LAT instrumental performance, the $\gamma$-ray background modeling, and the assumed dark matter density profile. For the analysis, the Fermi collaboration considered a variety of density profile models, and  found the resulting limits to be fairly insensitive to the inner slope of the DM density profile as long as such inner slope scales as $r^{-\gamma}$ with $\gamma< 1.2$. 
 
In recent years, a number of studies have advocated the presence of  black holes (BH) in dwarf galaxies in general, and in dwarf spheroidal (dSph) galaxies in particular. A sample of 151 dwarf galaxies exhibiting optical signatures of accretion by intermediate massive black holes (IMBH) was presented in Ref.~\cite{Reines:2013pia}. Such galaxies have masses of about $10^{7-9} M_{\odot}$ and velocity dispersion in the range $20-60\, {\rm km s^{-1}}$. More recently, a set of 28 active galactic nuclei (AGNs) were identified in nearby ($d < 80 Mpc$) low-mass, low-luminosity dwarf galaxies \cite{2014arXiv1408.4451M}. In both cases, the expected mass of the IMBH is lower than $10^6 \Msun$.This indicates that the process of BH formation in galactic nuclei could be very similar for regular and for small galaxies alike, contrary to what was previously thought.  There are also mild indications for the existence of BHs in dwarf spheroidal galaxies, such as NGC205, a satellite of the Andromeda galaxy, and in Ursa Minor and Fornax, satellites of the MW; The inferred mass of the BHs all have upper limits of $M_{\rm bh} \approx 10^{4} M_{\odot}$ \cite{Valluri:2005up,2013NewA...23..107N,2009ApJ...699L.113L,2012ApJ...746...89J}.  It is notable that the last two dwarfs are included in the Fermi analysis. Additional searches for IMBHs in dSphs are at present underway \cite{2013MmSAI..84..645N,Nucita:2013iu}. The black hole in $\omega$-Centauri \cite{Noyola:2008kt},  if it is conclusively found to be the remnant of a tidally disrupted galaxy \cite{Ideta:2004qz} instead of a globular cluster \cite{2012ApJ...751....6D}, can be added to the list.  This mounting observational evidence motivates us to extrapolate known correlations between black hole mass and velocity dispersion, $M_{\rm bh}-\sigma_{*}$,  or luminosity, of the host galaxy, to dwarf and dwarf spheroidal galaxies, and to study the effect of such compact objects on the dwarfs' galactic dark matter density profiles.  

The abundance of IMBHs in Milky Way-like halos was recently studied by means of numerical simulations in Ref.~\cite{Rashkov:2013uua}. This study suggests that about $70-2000$ IMBHs could be present in the sub-halo satellites of the MW, depending on the velocity dispersion of the stellar component that populates these sub-halos. One cannot however directly rely on this result to postulate the presence of IMBHs in the dwarf spheroidal galaxies as we are considering here, since those simulations do not correspond exactly with the observed local structure. Nonetheless, the results of numerical simulations inform us of how likely it might be that these objects are present in the local volume, and they motivate the study of how IMBH affect our inferences from a given Indirect detection observable. The contribution to the isotropic gamma ray background from annihilation of DM in spikes around supermassive black holes has been addressed recently in \cite{2014PhRvD..89d3520B}.

In this paper we address the effect of the presence of central black holes on constraints on dark matter pair annihilation from the non-observation of gamma-ray emission from local dwarf spheroidal galaxies. Notice that work along these same lines, prior to the Fermi LAT era, was presented by one of us and Collaborators in Ref.~\cite{Colafrancesco:2006he}, for the specific case of Draco. 

 \section{ The $\gamma$-ray signal from dark matter annihilation in the presence of a central Black Hole}
If dark matter is composed by WIMP-like particles, such particles generically have a small but not negligible probability to annihilate into Standard Model particles, which could eventually be detected. The ingredients that fix the flux of particles produced by this process are: first, the particle physics model, including the thermally averaged, zero-temperature pair-annihilation cross section times relative velocity,  $\langle\sigma v\rangle$,  the dark matter particle mass, $m_{\chi}$, and the the energy spectrum of the photons produced in each annihilation event,  $dN_{\gamma}/dE_{\gamma}$; second, the number density of WIMP pairs available to annihilate, which is proportional to the square of the dark matter mass density, $\rho(r)$, for the target under consideration, dwarf spheroidal galaxies in this particular case. Given these components, the integrated $\gamma-$ray signal flux, $\phi ( {\rm ph\,cm^2 s^{-1}})$, expected from the dark matter annihilation in the density distribution, $\rho(r)$, from a given angular region $\Delta\Omega$, is given by:

\begin{equation}
\phi(\Delta \Omega)=\frac{1}{4\pi}\frac{\sigmav}{2 m_{\chi}} \left(\int_{E_{\rm min}}^{E_{\rm max}}{\frac{dN_{\gamma}}{dE_{\gamma}}}dE_{\gamma}\right)\; J_{\Delta \Omega}
\label{eq:AnnFlux}
\end{equation}
with 
\begin{equation}
 J_{\Delta \Omega}=\int_{\Delta \Omega} {\int_{los} \rho^2(r(l,\theta),\sigmav,m_{\chi})\,dl d\Omega}.
 \label{eq:Jfactor}
\end{equation}

The last term, commonly referred to as the ``$J$-factor'', is the line-of-sight integral through the dark matter distribution integrated over the solid angle $\Delta \Omega$. Here the coordinate $r$, centered on the dwarf galaxy reads $$r(l,\theta) = \left(d^2+ l^2  - 2\,d\,l\,\cos(\theta)\right)^{1/2};$$ in the expression above, $d$ is the distance to the dwarf galaxy, and $\theta$ is the aperture angle between the line of sight direction and the axis connecting the Earth to the galaxy's center. We will integrate the $J$-factor within an angular radius of $0.5\deg$, i.e. $\Delta \Omega=2.4 \times 10^{-4} {\rm sr}$, following the Fermi analysis \cite{Ackermann:2013yva}. 

 In what follows, we  will use the constraints derived from the Fermi analysis, Ref.~\cite{Ackermann:2013yva}, and we will thus only consider specific annihilation final states. Note that in standard analysis the $J$-factor does not depend on the cross-section, nor on the mass of the DM particle. In our case, however, the presence of the black hole will enforce such dependence, as we explain below, and we thus feature such dependence explicitly in Eq.~(\ref{eq:Jfactor}). 

\subsection{The DM density profile}
The inner slope of the DM density profile in dwarf galaxies remains a topic of intense debate \cite{2007ApJ...657..773V,walker2009,walker2011,Jardel:2012am}. The Fermi collaboration  \cite{Ackermann:2013yva} used both, a  cuspy (Navarro, Frenk and White, or NFW~\cite{Navarro:1996gj})  and a cored (Burkert~\cite{Burkert:1995yz}) density profiles in their analysis. The resulting $J$-factors, however, are almost insensitive to the choice of density profile, as long as the inner slope $r^{-\gamma}$ has $\gamma< 1.2$ \cite{Ackermann:2013yva}. This will no longer be true when we account for the presence of a black hole, primarily because the typical resulting slope in the density profile after adiabatic accretion exceeds $\gamma\gg1.2$. In our analysis we use as a starting configuration for the DM density profile of a given dwarf the same benchmark density profiles considered in Ref.~\cite{Ackermann:2013yva}, i.e. the NFW and the Burkert density profiles, respectively defined as:

\begin{eqnarray}
\rho_{\rm NFW}&=&  \frac{\rho_s}{r/r_s (1+r/r_s)};\\
\rho_{\rm Burk}&=& \frac{\rho_{0}}{(1+r/r_0) (1+r^2/r_0^2)},\
\label{eq:profiles}
\end{eqnarray}
where $\rho_s$ and $r_s$ are the characteristic density and scale radius, and $\rho_0$ and $r_0$, the central density and core radius.  We use specific values for the two parameters for each of the dwarf galaxies in the analysis, as inferred from the maximum velocity, and the mass contained up to the radius of maximum velocity listed in  Table 2 of Ref.~\cite{Martinez:2013els}. For these parameters, we have verified that the $J$-factor, as defined in Eq.~(\ref{eq:Jfactor}) corresponds to that reported by the Fermi collaboration, listed in columns 5 and 6 of table \ref{tab:dsphs}, within the error bar, for each of the dwarfs. 

To take into account how the DM density profiles are affected by the presence of a black hole of a given mass, and consequently how such black holes affect the expected annihilation signal, we make the hypothesis that a black hole of a given mass, $m_{\rm bh}$, formed adiabatically at the center of each galaxy. In this process, it is expected that the inner dark matter halo density profile is modified due to adiabatic contraction, evolving from an initial density profile scaling as $\rho_i(r)= r^{\gamma}$  into a final profile of the form $\rho_f (r) \propto r^A$, with $A = (9-2\gamma)/(4-\gamma)$, for $0<\gamma <2$ \cite{Quinlan:1994ed,Gondolo:1999ef,Ullio:2001fb}.

The relation between the initial and final slope due to the adiabatic contraction process is obtained under the assumption that all particles are in circular orbits. For simplicity, we use this same approximation for both the NFW and the cored initial density profile \footnote{It was shown that this relation does not hold for a cored profile, $\gamma=0$, if it satisfies the condition of being an analytic core \cite{Quinlan:1994ed}. However, since the Burkert profile does not fulfill such condition, the final slope will be somewhere in the range between $2<\gamma_{sp}\leq 2.25$, where the upper limit is obtained when using the circular orbit approximation. We rely on this approximation since this will give us the maximum effect that the black hole could produce. As we show below, this choice will not affect significantly our conclusions regarding cored profiles anyways.}. Summarizing, the inner density profile of the dwarfs hosting a black hole is considered to have a inner slope of $A=7/3$ for the NFW profile, and $A=9/4$ for the Burkert profile.  
To calculate the final density profile, we assume an initial dark matter distribution $\rho_i(r)$ made of particles that are all on circular orbits. If a black hole grows adiabatically at the center of this distribution, the angular momentum of each particle remains invariant, since the central black hole exerts no torque on any dark matter particle. This implies that 
\begin{equation}
r_i M_i(r_i) = r (M (r )+M_{\rm bh}),
\label{eq:adcontra1}
\end{equation}
where $M_i(r_i)=4\pi\int \rho_i(r) r^2 dr$ is the dark matter mass enclosed within radius $r_i$  initially, and  $M(r)$ is the dark matter mass enclosed within radius $r$, to which the particles are displaced after the growth of the black hole. Conservation of dark matter mass implies $M_i(r_i)=M_f(r_f)$.  The density profile modified by the adiabatic growth of the black hole is then calculated iteratively using Eq.~\eqref{eq:adcontra1} and:
\begin{eqnarray}
\rho(r)&=&M'(r)/4\pi r^2, \quad \mbox{with}\nonumber\\
M'(r)&=&\frac{M(r)+M_{\rm bh}}{r_i-r+M_i(r_i)/M'_i(r_i)}.
\label{eq:adcontra2}
\end{eqnarray}
For a completely self-consistent approach one would need to refer to the full phase space distribution function for the DM density profile, and implement the appropriate adiabatic invariant. However, the error estimating the new density profile with the circular orbit approximation is smaller than the uncertainty in the normalization of the initial density profile (see e.g.~\cite{Ullio:2001fb} and a more recent review of this calculation including general relativistic corrections given in Ref.~\cite{Sadeghian:2013laa}). We also point out that the DM density profile inner slope is not affected by the circular orbit assumption. Hence we adopt the simplest approximation because the time required to compute the new density profile is highly reduced, and the calculation is less susceptible to numerical errors. This is especially convenient since we will need to evaluate the new density profile for each of the dwarf galaxies several times.

Our analysis procedure is then as follows: For each of the dwarf galaxies considered in Ref.~ \cite{Ackermann:2013yva} we will: 

(1) Pick an initial halo model, as defined in Eq.~(\ref{eq:profiles}), with parameters consistent with the observed velocity dispersion for each dwarf spheroidal galaxy. We carry this out for both initial density profiles choices (NFW and Burkert).  

(2) Assign to each dwarf galaxy a black hole with a mass also consistent with the observed velocity dispersion, or luminosity, as we discuss next, and calculate the resulting modification to the density profile due to the adiabatic contraction, according to Equations \eqref{eq:adcontra1} and \eqref{eq:adcontra2}. 

(3) Calculate the corresponding $J$-factor, from Eq.~\eqref{eq:Jfactor}.

(4) Derive the resulting constraints on the DM particle properties parameter space defined by ($m_\chi, \langle \sigma v\rangle$).

\subsection{The black hole mass}
\begin{figure*}[t!]
\centering

\includegraphics[width=0.47\textwidth]{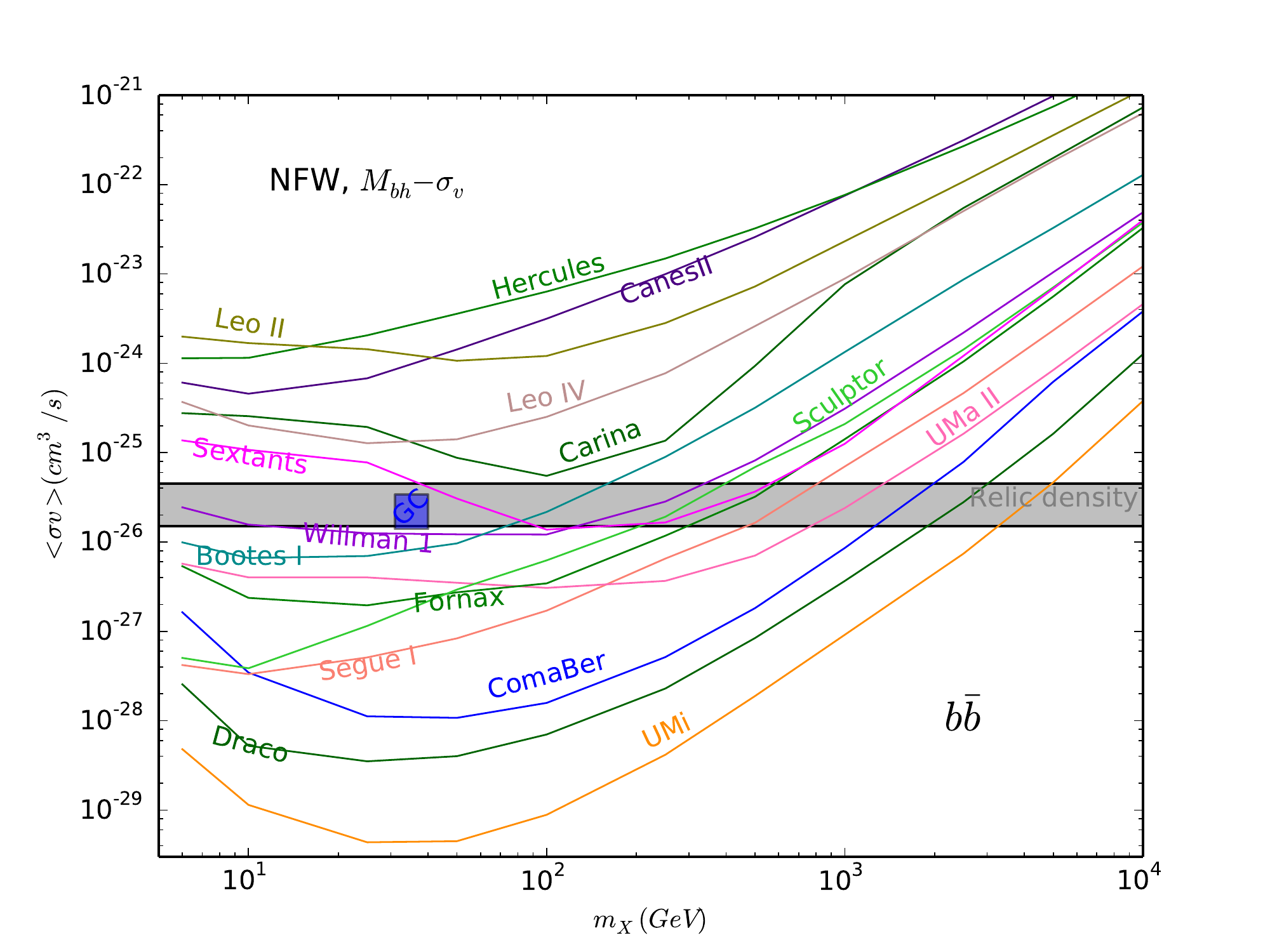}
\includegraphics[width=0.47\textwidth]{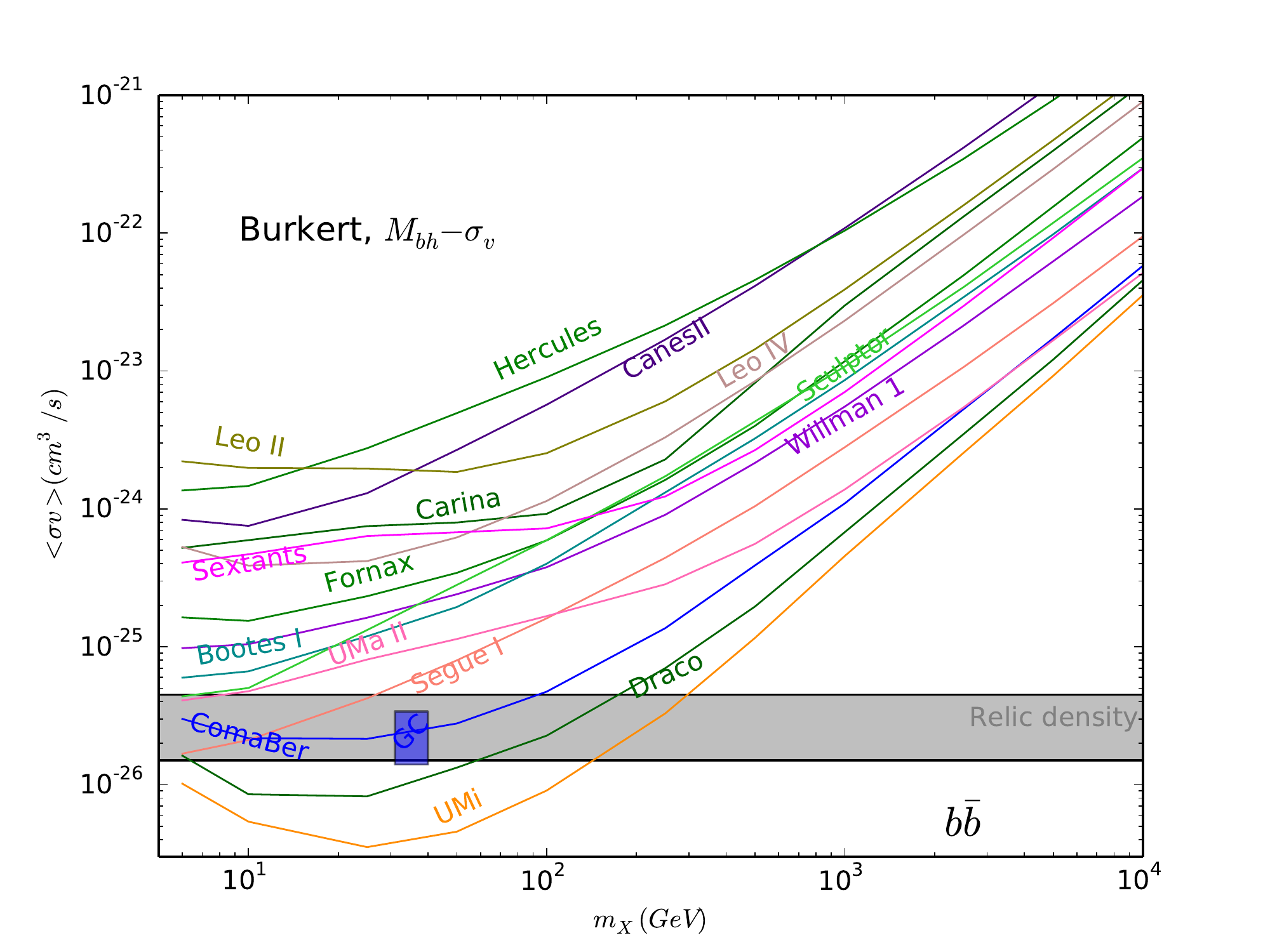}
 \vspace{-15pt}
  \caption{DM annihilation cross-section constraints for the $b\bar{b}$ final state, for individual dSph, assuming an initial NFW DM density distribution (left), and a Burkert profile (right). In both cases, BHs were assigned using the Tremaine relation, $M_{\rm bh}-\sigma_{v}$ \cite{Tremaine:2002js}. } 

\label{fig:Fermi1}
\end{figure*}

As we described in the Introduction, recent observational evidence for black holes in dwarf galaxies suggests that the known relations  for the  black hole mass and stellar velocity dispersion, or luminosity, of the host, for massive galaxies could also hold for the smallest known galaxies, dwarf and dwarf spheroidal galaxies. We will extrapolate some of these relations to the lower end mass of galaxies with the aim of studying how the presence of a black hole could affect our inferences on the dark matter particle physics.

Different relations between the mass of the black hole and the structural parameters of the host galaxy have been proposed, see for instance Ref.~\cite{Magorrian:1997hw,2000ApJ...539L...9F,Tremaine:2002js,2011ApJ...737...50V,2011MNRAS.412.2211G,McConnell:2012hz}. For the present analysis we consider three of the most widely used relations: First, the \emph{Magorrian} relation \cite{Magorrian:1997hw},
\begin{equation}
M_{\rm bh}=  0.0013 L_{*},  
\end{equation}
which relates the black hole mass, $M_{\rm bh}$, to the luminosity of the host galaxy $L_*$, for which we assumed a mass to light ratio of $M_{*}/L{_*}=1$ for the stellar distribution. The actual $M/L$ ratio may vary by about a factor of two depending on the details of the stellar populations, see \citet{2013pss5.book.1039W} and references therein. This was the  first proposed relation of this kind. Afterwards, many different studies have proposed variants to that, although the main idea is that the mass of the black hole is driven by the the properties of the luminous component, mainly the galactic bulge. Other studies argue for a correlation with the dark matter halo (see e.g. Ref.~\cite{2011ApJ...737...50V}).

Our second relation to infer the BH mass is the \emph{Tremaine} relation \cite{Tremaine:2002js},
\begin{equation}
{M_{\rm bh} \over \Msun}= 
\begin{cases} 10^{6.91} \left(\frac{\sigma_*}{100\,{\rm km/s}}\right)^4  & (\sigma_*\ge 6\,{\rm km s}) 
\\ 
100 & (\sigma_*< 6\,{\rm km/s}). \end{cases}
\label{msigma}
\end{equation}
 between the black hole mass and the velocity dispersion, $\sigma_*$, of the stellar component. 
 
Finally, we also consider a more recent proposal, the {\em McConell \& Ma} relation \cite{McConnell:2012hz},
\begin{equation}
{M_{\rm bh} \over \Msun}= \begin{cases} 10^{8.32} \left(\frac{\sigma_*}{200 \, {\rm km s^{-1}}}\right)^{5.64}  & (\sigma_*\ge 15\,{\rm km s}) 
\\ 
100 & (\sigma_*< 15\,{\rm km/s}). \end{cases}
\label{eq:McConell}
\end{equation} 
which also relates the black hole mass to the velocity dispersion. This last relation is the steepest one among the prescriptions we study here.  In general, different samples of galaxies and observational techniques result in steeper, or less steep, relations. However, there is some consensus that the mass of the black hole correlates better with the velocity dispersion. In particular, the Tremaine relation produces predictions in between the Magorrian and the McConell \& Ma relations, and therefore we will use it as our benchmark case. We will use the other two relations to bracket an upper (Magorrian) and lower (McConell \& Ma) limit to the effect under consideration. An additional assumption for the three relations, is the presence of a minimum mass of the black hole set to $M_{min}=100\ \Msun$, to resemble  black hole formation as Population III remnants. 

In columns 7 and 8 of Table ~\ref{tab:dsphs} we list the masses of the black holes corresponding to each of the dwarf galaxies under consideration, along with the galaxies' distance, luminosity, mean velocity dispersion, and with the $J$ factors for a NFW and a Burkert profile, for the Tremaine (column 7) and for the Magorrian relation (column 8). As for the McConell \& Ma relation, all the galaxies are seeded with a black hole corresponding to the minimum mass of 100 $\Msun$.

\section{Fermi limits from dwarf spheroidal galaxies in the presence of a central Black hole}

\begin{figure*}[t!]
\centering
 
\includegraphics[width=0.47\textwidth]{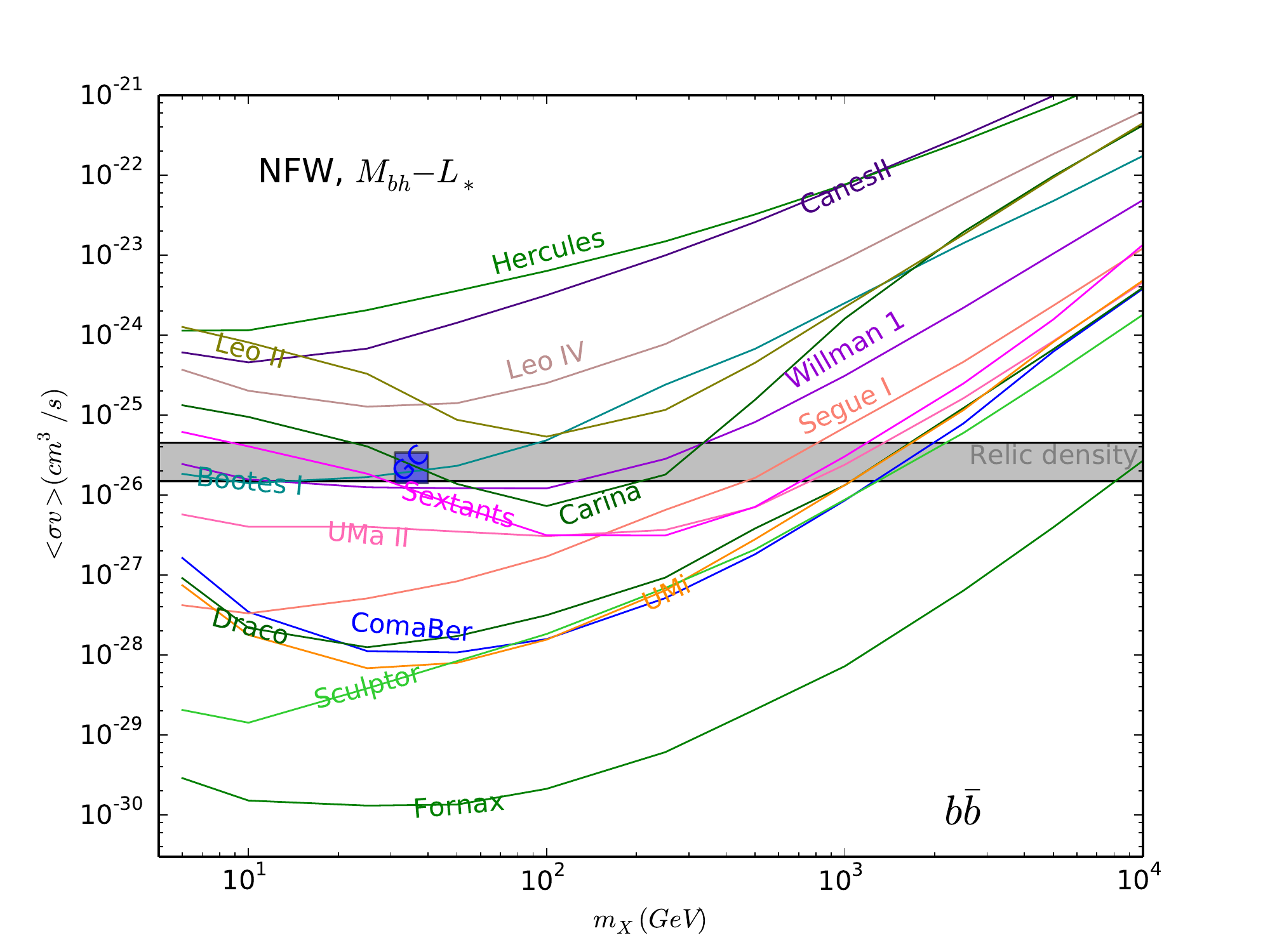}\includegraphics[width=0.47\textwidth]{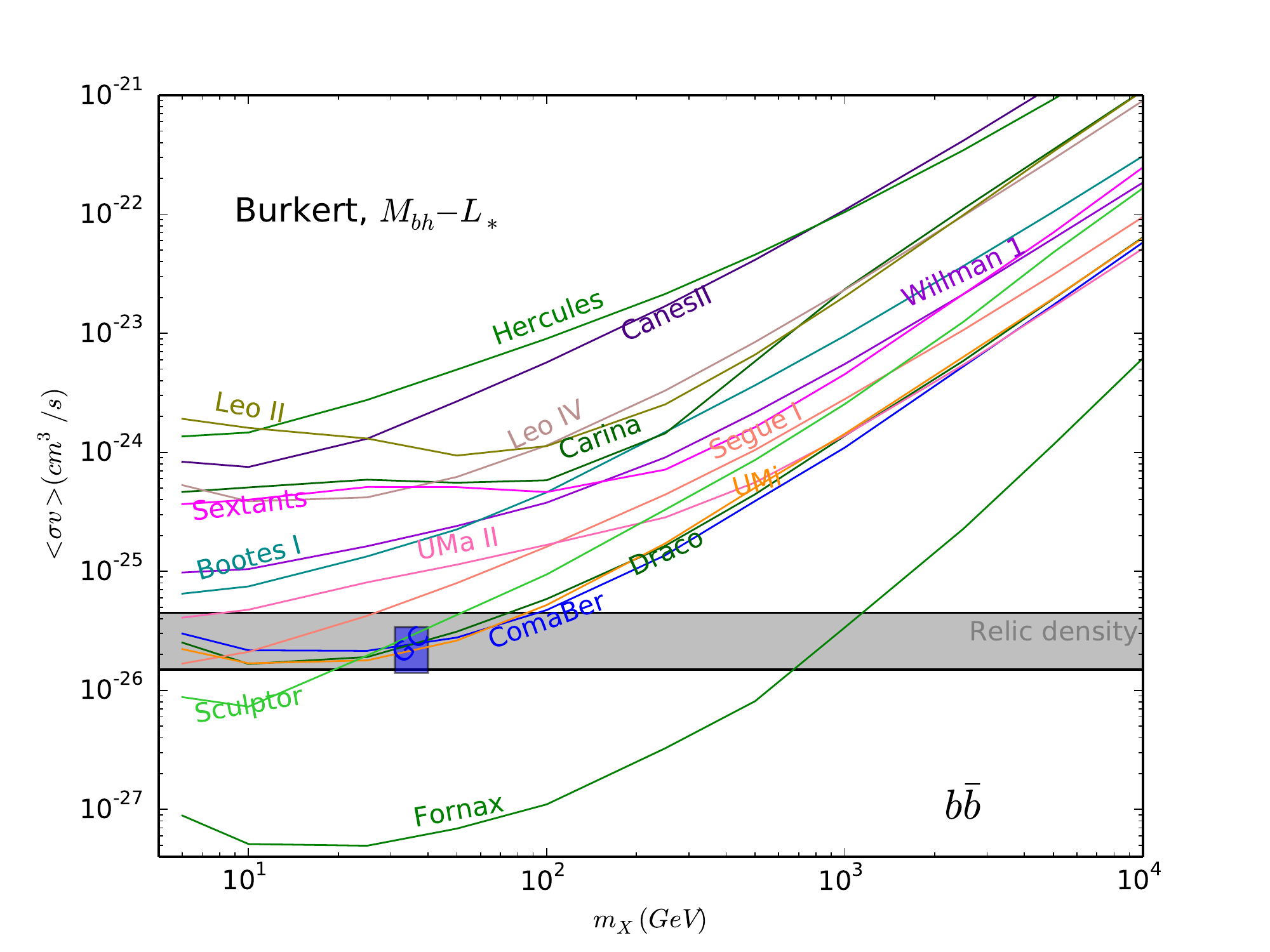}

 \vspace{-15pt}
  \caption{As in figure \ref{fig:Fermi1}, but for the Magorrian relation, $M_{\rm bh}-L_{*}$ for the mass of the black hole \cite{Magorrian:1997hw}.} 
\label{fig:Fermi2}
\end{figure*} 

Given a prescription for the central black hole mass, a choice for the initial dark matter density profile, and, finally, the adiabatic contraction recipe we described in the previous section, we calculated the new dark matter density profile and the corresponding $J-$factor for each of the dwarf galaxies. The new density profiles are usually singular in the $r\to0$ limit, in such a way as to produce an a priori divergent $J$ factor. Physically, such large densities lead to high annihilation rates, and an ensuing inner cut-off radius $r_{\rm cut}$ naturally appears, within which the DM density becomes constant, due to the balance between the annihilation rate and the gravitational infall rate of DM particles due to the presence of the BH. The DM annihilation rate is given by $\dot{\rho}=-\sigmav \rho^2$,

\begin{equation}
\rho(r,t)=\frac{\rho(r,t)}{1+\rho(r,t_f)\sigmav (t-t_f)}
\end{equation}
where, $t$ is the present time, and $t_f$ is the formation time of the BH. The maximum density is reached if  $(t-t_f)<<(\sigmav n(r,t_f))^{-1}$. The value of this maximum density, considering a  reasonable time formation of the BH in dSphs ($\sim 10{\rm Gyr}$), takes the typical value:

\begin{equation}\label{eq:rhocutoff}
\rho_{{\rm max}}=3 \times 10^{18} \left(\frac{m_{\chi}}{100 {\rm GeV}} \right) \left(\frac{10^{-26} {\rm cm^3 s^{-1}}}{\sigmav} \right) \Msun \rm kpc^{-3}.
\end{equation}

Notice that in principle this maximum density depends  on the formation time of the BH in each galaxy, which we are now approximating to be the same for all the dSphs. We define a cut-off radius $r_{\rm cut}$ as the radius at which the density profile is equal to the maximum density quoted in Eq.~(\ref{eq:rhocutoff}). Capture of DM particles by the BHs could also be taken into account, however the Schwarzschild radius of the BHs is in general smaller than  $r_{\rm cut}$. 

Now,  $J-$factor is computed by adding the contribution of a sphere of constant density $\rho_{\rm max}$ and radius $r_{\rm cut}$, with the usual integral of the density profile from $r_{\rm cut}$ outwards. It is clear now where the dependence of the $J-$factor with the annihilation cross section and the mass of the particle comes from.   
 
 It is worth recalling that the density profile resulting after adiabatic contraction is still consistent with current velocity dispersion data of each dSph. The effect of the contraction affects the density profile on scales below  $100 {\rm pc}$, the typical scale for which data is available. Of course, this is only valid as long as the mass of the IMBH is much smaller than the contained dark matter mass within the half mass radius. The validity of this condition can be assessed by using a simplified Jeans equation to estimate the mass contained within a radius $r_{\rm half}$, given a velocity dispersion profile, $\sigma$, as was done in  \cite{walker2009}. 
\begin{equation}
M(r_{\rm half}) =\mu r_{\rm half} \sigma^2 ,
\end{equation}
where $\mu$ is just the numerical factor arising from the constants involved. Note that this estimate of the dark matter mass contained up to $r_{half}$ is in good agreement with those obtained from a full Jeans/Markov Chain Monte Carlo analysis. In the derivation of the expression above there is no information a priori of the form of the dark matter density profile, then it is also valid for the profile after adiabatic contraction, and with the IMBH:
\begin{equation}
M_{\rm DM}(r_{\rm half})+M_{\rm bh}=\mu r_{\rm half}\sigma_{\rm DM+BH}^2 
\end{equation}.
Combining the equations above, and using the  fact that adiabatic contraction of the halo conserves mass, we have that 
\begin{equation}	
\frac{\sigma_{\rm DM+BH}^2}{\sigma^2}\approx 1+\frac{M_{\rm BH}}{M(r_{\rm half})}.
\end{equation}

Therefore as long as the mass of the BH is much smaller than the mass within $r_{\rm half}$, of the order $\lesssim10^7 \Msun$, the presence of the black hole will still be consistent with current velocity dispersion data. The condition outlined above is always well satisfied by the IMBHs mass under consideration here.

We now utilize the new calculated results for the $J$ factors in the presence of a central black hole to find the corresponding constraints to the dark matter particle properties.  We first consider each of the dwarf galaxies individually.  Fermi LAT observations  give us the actual observed flux of gamma rays coming from a certain direction, once the known sources and backgrounds have been subtracted. The fact that none of the known dwarf galaxies has been detected over background provides a maximum flux of gamma rays in a certain energy bin, that is, it gives us an upper limit for the left side of Eq.~(\ref{eq:AnnFlux}), $\phi$. Given a $J$ factor for a given galaxy, i.e. the third term in the right side of the same equation, we can read an upper limit for the cross section $\langle\sigma v\rangle$ as a function of the mass of the particle, $m_{\chi}$, given the $\gamma$-ray yield $dN/dE$ from a particular model. This is essentially what was done by the Fermi Collaboration to derive the colored lines of figure 1 in Ref.~\cite{Ackermann:2011wa}.  

Notice that the upper limit on the flux of gamma rays from dark matter annihilation set by the Fermi Collaboration employs a $J-$ factor that is independent of the particle mass and pair-annihilation cross section. This no longer holds for the present case, given the cutoff effect described above, which makes $J_{bh}=J_{bh}\left(\sigmav_{bh}, m_{\chi} \right)$. We can however directly utilize the Fermi Collaboration constraints on $\sigmav$ from Ref.~\cite{Ackermann:2011wa}, for a given annihilation pathway, as follows:
\begin{equation}
\sigmav J = \sigmav_{bh} J_{bh}\left(\sigmav_{bh}, m_{\chi} \right)   \quad \forall \quad  m_{\chi}.
\label{eq:newconstr}
\end{equation}

We use the constraints from individual galaxies derived by the Fermi Collaboration, see Table 2-7 in \cite{Ackermann:2013yva}, to compute the corresponding constraints when including the effects of a central black hole in such galaxies, by solving for $\sigmav_{bh}$ in Eq.~\eqref{eq:newconstr}, together with Eq. \eqref{eq:Jfactor}. 
\begin{figure*}[t!]
\centering
 
\includegraphics[width=0.47\textwidth]{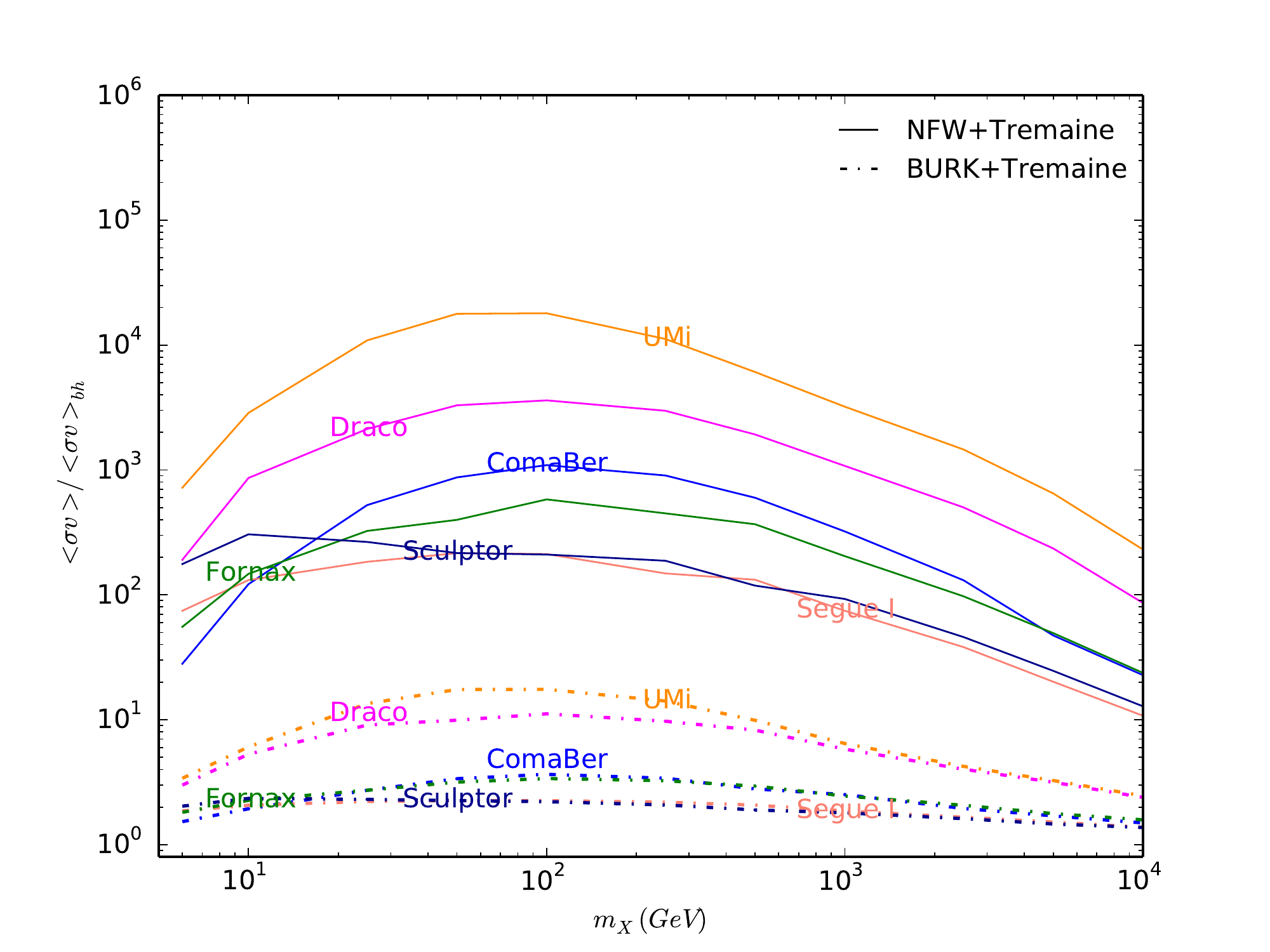}
\includegraphics[width=0.47\textwidth]{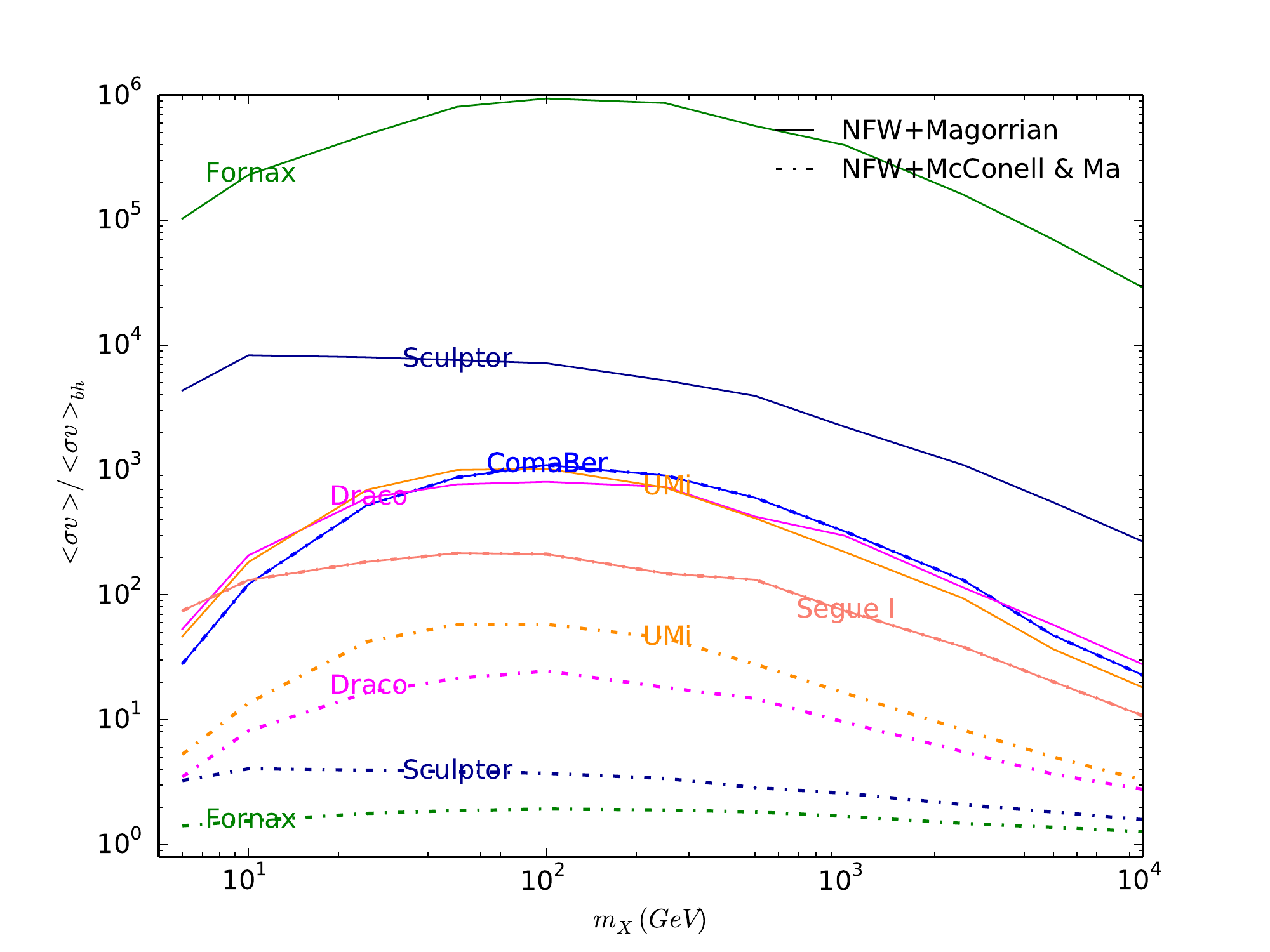}
 \vspace{-15pt}
  \caption{Enhacement factor, $\sigmav/\sigmav_{bh}$, for the DM annihilation cross-section constraints for different black hole mass scenarios, for selected  individual dSph. (Left)We compare the  NFW and Burkert profiles both assuming a Tremaine relation for the black hole mass. (Right) For the NFW we compare the Magorrian and the minimum mass black hole scenario.} 

\label{fig:Enhance}
\end{figure*} 

We show our results, for annihilation in the $b\bar{b}$ channel, in Figs. \ref{fig:Fermi1} and \ref{fig:Fermi2}.  Fig.\ref{fig:Fermi1}  shows the individual constraints for all dwarf galaxies considered in the Fermi Collaboration analysis, assuming the Tremaine relation, $M_{\rm bh}-\sigma_v$, for a NFW profile (left panel) and for the cored Burkert profile (right panel). We note that for Draco and, especially, for Ursa Minor the effect of a central black hole is dramatic, and leads to constraints, for the particular $b\bar{b}$ annihilation pathway, of up to almost four orders of magnitude stronger than the classic pair-annihilation cross section value invoked for thermal decoupling production of DM in the early universe (for s-wave annihilation), $\sigmav\simeq3\times 10^{-26}$ cm$^3$/s (we shade in grey a representative range for $\sigmav$ around that central value). In the 10-100 GeV mass range, 7 individual galaxies produce constraints that rule out such a cross section for some value of the particle dark matter mass. The results for the Burkert profile are relatively less dramatic; constraints from the Ursa Minor dSph in this case still rule out an s-wave thermally produced relic in the 10-100 GeV range. It is important to emphasize that unlike in the case where no black hole is present, the seed DM density profile produces important effects when adiabatic contraction onto a central massive object is in place.

Fig. \ref{fig:Fermi2} mirrors what shown in Fig.~\ref{fig:Fermi1}, but now with the assumption of the Magorrian relation, i.e. $M_{\rm bh}-L_{*}$. In this case the constraints we obtain are significantly stronger when the initial density profile is cuspy than when it is cored. On the other hand, constraints using the Magorrian relation tend to be stronger than with Tremaine relation for some of the galaxies and not for others, changing the hierarchy of constraints. For example, when using the Tremaine relation the  Draco and UMi galaxies are the most constraining, while for the Magorrian relation the more luminous Fornax galaxy is expected to host a comparatively much larger black hole, producing the most stringent limits. Notice that while in both cases Fornax is the galaxy that hosts the most massive BH, it is only  when we use the Magorrian relation that the effect of the BH dominates over the fact that UMi and Draco are significantly closer than Fornax. Taken at face value, the Fornax limits for an NFW profile rule out thermally produced ``vanilla'' relics for masses up to about 10 TeV, and thus close to the unitarity limit \cite{Griest:1989wd}.

We illustrate in figure \ref{fig:Enhance} the relative enhacement factor $\sigmav/\sigmav_{bh}$ for the different combinations of density profile and black hole mass relation, and for selected galaxies, those that are more relevant for the constraints.  Maximal enhancements are typically obtained for particle masses in the 100 GeV range. The peculiar functional form shown in the figure results from a combination of two contributions to the $J$-factor: on the one hand, the one from the inner sphere $r<r_{\rm cut}$, and on the other hand that from the density profile for $r>r_{\rm cut}$. Since the maximal density scales linearly with mass, the maximal {\em number} density is mass-independent. The cutoff radius, however, grows for smaller masses. The competing contributions are then compounded via Eq.~\eqref{eq:newconstr} in a non-trivial way, resulting in the mass-dependent enhancements shown in the figure.

It is also remarkable that in the presence of a central black hole, Fermi limits from dSph would conclusively rule out the possibility that the Galactic center excess originate from dark matter annihilation, as entertained in Ref.~ \cite{Hooper:2011ti}. We highlight the favored mass and pair-annihilation region for the Galactic center excess with a blue rectangle in all plots shown in Fig.~\ref{fig:Fermi1} and \ref{fig:Fermi2}.
 
 The above results rely on two important assumptions: (i) that the black hole formed adiabatically, and (ii) that it was formed at the galaxy's dynamical center. If either one or the other assumption is not valid, the effect of the enhancement on the $J-$factor, and therefore on the limit to the cross section, would be  smaller than what we derive here. For instance, for the NFW initial profile,  if the black hole growth were instantaneous, the slope of the density profile would be $\gamma=4/3$ \cite{Ullio:2001fb}, smaller than the adiabatic growth ( $\gamma=7/4$), but still larger of what is covered by current FERMI analysis ( $\gamma=1.2$). On the other hand, if the BH was formed off-center, two factors  compete to determine the final density profile: the BH needs to have enough time to spiral in from the site in which it was formed, and the initial BH seed mass is small enough to adiabatically grow once it is on center. The lower the initial BH seed mass, the closer to the center it must form in order to have enough time to spiral in and continue growing adiabatically. If the initial BH seed mass is of the order of the final one, the process of in-spiral would  expel more dark matter particles than those that would be  attracted by the adiabatic growth later on, in which case the final density would be even lower than it was initially. The details of the BH formation can be different for each of dSphs, and given that there are not clear detection of IMBHs in this galaxies it is not possible to account for those effects described above without making more assumptions.

\subsection{Joint analysis for the 15 dSphs}

In addition to individual objects, the Fermi Collaboration produced limits on the gamma-ray flux, and thus on the dark matter annihilation rate, from a combined analysis of 15 dwarf galaxies sample \cite{Ackermann:2011wa}. Under the assumption that DM properties, annihilation cross section, mass, etc, are the same in all the dwarf galaxies, Ref.~\cite{Ackermann:2011wa} exploited the fact that the sensitivity to a weak signal is potentially increased if multiple signals are combined. To do so, Ref.~\cite{Ackermann:2011wa} created a joint likelihood function from the product of the 15 individual likelihood functions and performed a maximum likelihood analysis. To reproduce such a procedure taking into account the presence of a black hole is out of the scope of this paper, and would be very computationally costly because of the dependence of the $J$ factors on masses and annihilation rates.   However, we attempt here to obtain an estimate of a combined constraint making simple assumptions and utilizing individual limits from all dwarfs.

The constraints from the Fermi observations mainly arise from the lack of photon counts in addition  to those expected from  known background sources. Given an exposure, i.e. the product of an effective Area, $A_{\rm eff}$ and an observing time $T_{\rm obs}$, the photon number counts is roughly $N_{\gamma}\sim A_{\rm eff} \, T_{\rm obs} \,\phi_{\gamma}$. Assuming Poissonian fluctuations, the total number of photons detected from dwarf $i$ is  $N_{i} \pm\sqrt{N_{i}}$. Given that no significant excess of gamma rays from the directions of the dwarfs had been detected, the limit on the cross section can be obtained from
\begin{equation}
\sqrt{N_{i}}\propto \frac{\sigmav_i}{m_{\chi}^2} J_{\Delta \Omega,i}.
\label{eq:error}
\end{equation}
Considering now all $n$ dwarfs, the total (background) photon count, assuming no correlation in the photon counts from different dwarfs, is $N=\sum_i N_i\pm\sqrt{\sum_i N_i}$.

As a result, the limit $\sigmav$ from the combined observation all dwarfs, from Eq.~(\ref{eq:error}) with $N_i=N$, gives
\begin{equation}
\sigmav=\frac{\sqrt{\sum_i \left(\sigmav_i J_{i}\right)^2}}{\sum_i J_i} \quad \forall \ m_{\chi}.
\label{eq:combined1}
\end{equation}
This relation holds whether we take into account the presence of the black hole or not. In particular when we consider the presence of black hole, the $J-$ factors depend on the cross section, and the mass of the particle,  i.e.
\begin{equation}
\sigmav=\frac{\sqrt{\sum_i \left(\sigmav_i J_{i}\left(\sigmav_i,m_{\chi}\right)
 \right)^2}}{\sum_i J_i\left(\sigmav_i,m_{\chi}\right)} \quad \forall \ m_{\chi}.
\label{eq:combined2}
\end{equation}

Our simple ``stacking'' method does not imply that the global limit is always better than the individual limits. Just to illustrate this suppose that all dwarfs have the same $J_i=J$. In this case, from Eq.~\eqref{eq:combined1}, the limit improves to $\langle\sigma v\rangle = \frac{\sqrt{\sum_i \langle\sigma v\rangle_i^2}}{n}.$ Now, if all dwarfs have the same ``noise'', i.e. $N_i=\bar N$, then all limits are the same, $\langle\sigma v\rangle_i=\bar{\langle\sigma v\rangle}$ and the global limit
$\langle\sigma v\rangle = \frac{\bar{\langle\sigma v\rangle}}{\sqrt{n}}.$ Suppose now that one has two dwarfs, with the same $J$ factors, but one (dwarf 2) has a background count 4 times larger than the second, thus $\langle\sigma v\rangle_2=2\langle\sigma v\rangle_1$. In this case the global limit combining the two dwarfs is $\langle\sigma v\rangle = \langle\sigma v\rangle_1\frac{\sqrt{4+1}}{2}\simeq 1.12\langle\sigma v\rangle_1,$ hence from combining the two dwarfs one gets a limit which is 10\% worse than for dwarf 1 alone. Similarly, if a dwarf has a significantly lower $J$ factor, say by a factor 2, so $J_1=2 J_2$, and $N_1=N_2$, then the global limit is a factor $\sqrt{5}/2\simeq1.12$ worse than the limit from dwarf 1 alone.

When we apply this stacking procedure for the no-black hole case, we find that compared to the Fermi Collaboration joint likelihood analysis we produce an underestimate of a factor of 0.25 for low energies, and of a factor of 2 for large energies. This brackets the expected range of systematic uncertainty from utilizing the simple combined analysis approach outlined above versus a more sophisticated joint likelihood treatment.

In our results including the black holes some of the individual constraints turned out to be stronger than the combined one, as was expected from our stacking method; it is not unreasonable that one would find this to be the case even if a more complex stacking method is used, since with central black holes the differences in $J$ factors among dwarfs are very large.
\begin{figure}[t!]
\centering
\includegraphics[width=0.5\textwidth]{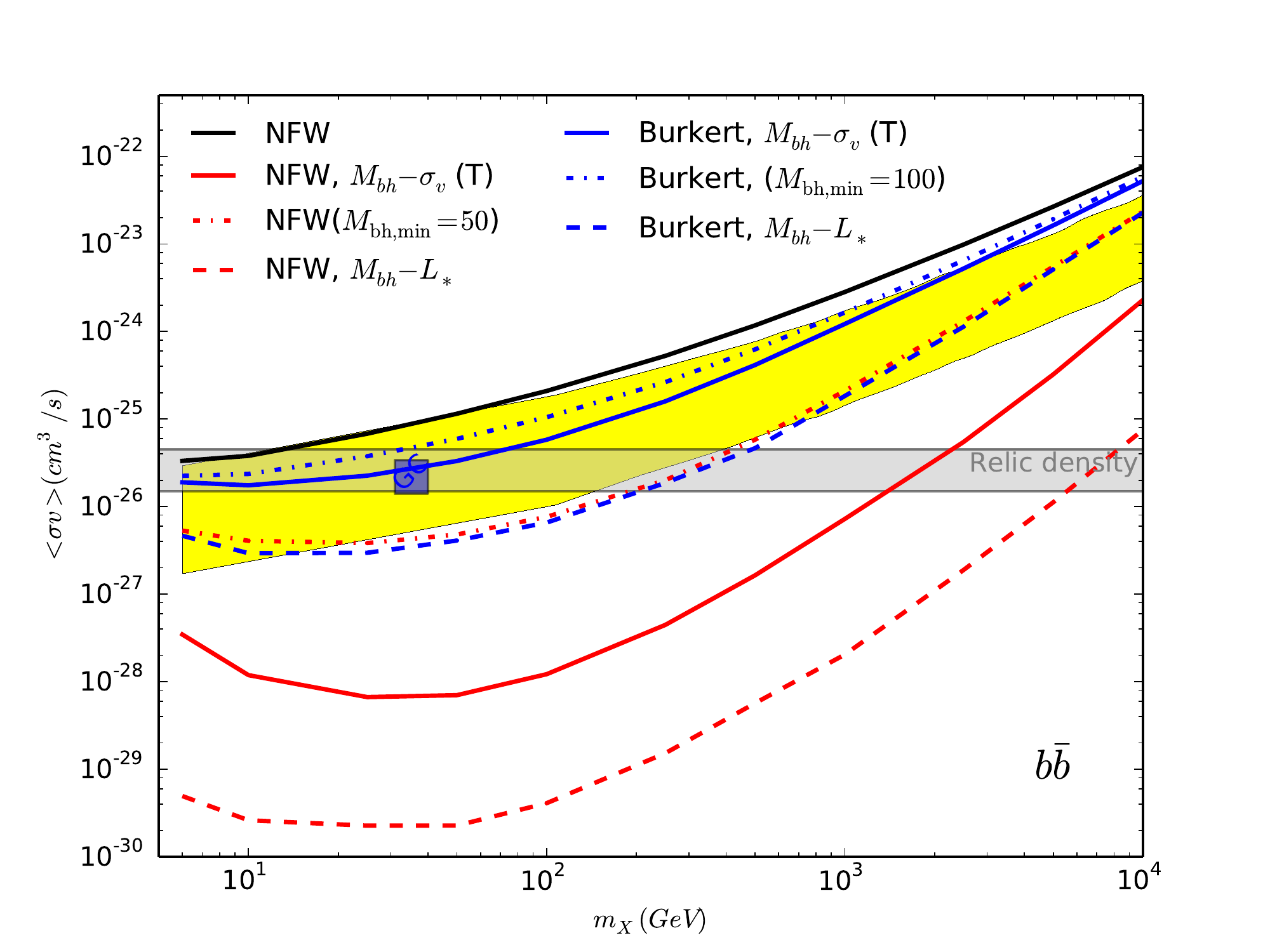}
 \vspace{-15pt}
  \caption{ DM annihilation cross-section constraints derived from a combined analysis of 15 dwarf spheroidal galaxies assuming an NFW (red) and Burkert (blue) density profile. We have assigned the black hole using the three different relations described in the text. The yellow band corresponds to the $95\%$ confidence level derived by Fermi Collaboration from their combined analysis of the same set of galaxies assuming an NFW dark matter distribution\cite{Ackermann:2013yva}}

\label{fig:FermiJoint}
\end{figure}  
A comparison of the combined constraints obtained in the three scenarios for the black hole assignment we consider here, NFW+Tremaine/Magorrian/McConnel\& Ma relation, is shown in Fig.~\ref{fig:FermiJoint} (red lines). In the figure we also show those same three scenarios but with a Burkert density profile (blue lines), and the constraint from the Fermi analysis (black line); in this case the combined constraints fall within $95\%$ confidence limit (yellow band) derived by Fermi Collaboration, from the combined analysis of the same set of galaxies, regardless of the black hole mass scenario.  

For the sake of illustration, in the figure we also show the case of a NFW profile for the 15 stacked dwarfs, all endowed with a black hole mass of $50\ \Msun$. In this case, we obtain combined constraints which approximately fall within the Fermi uncertainty band. Any minimal black hole mass larger that $50\ \Msun$ would lead to a significantly stronger combined limit. The figure also illustrates that the combination of NFW+ Magorrian relation gives the strongest limit. This scenario would be able to exclude all the thermal vanilla WIMP candidates lighter than $10 TeV$.
 
\begin{figure*}[t!]
\centering

\includegraphics[width=0.47\textwidth]{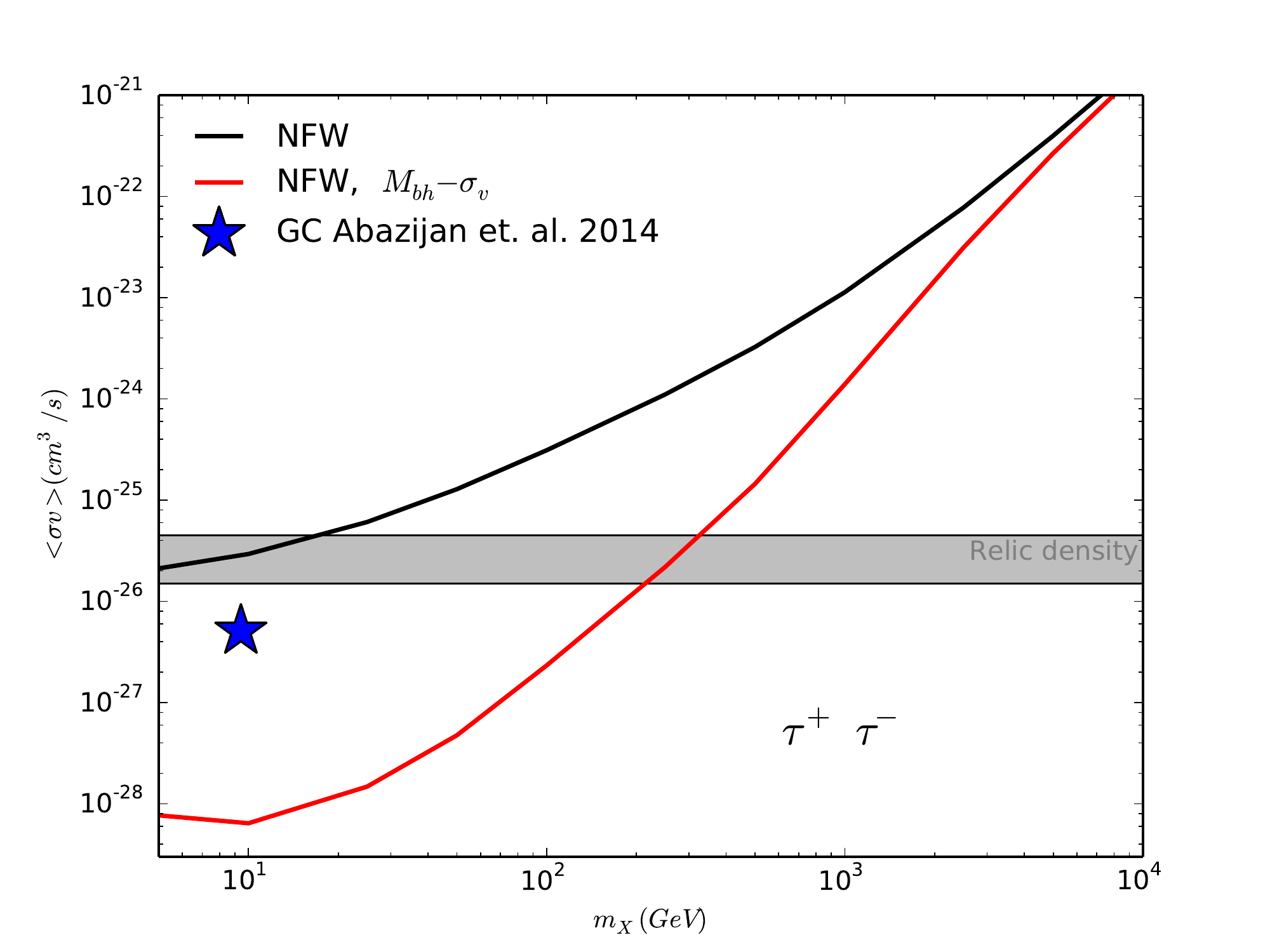}\includegraphics[width=0.47\textwidth]{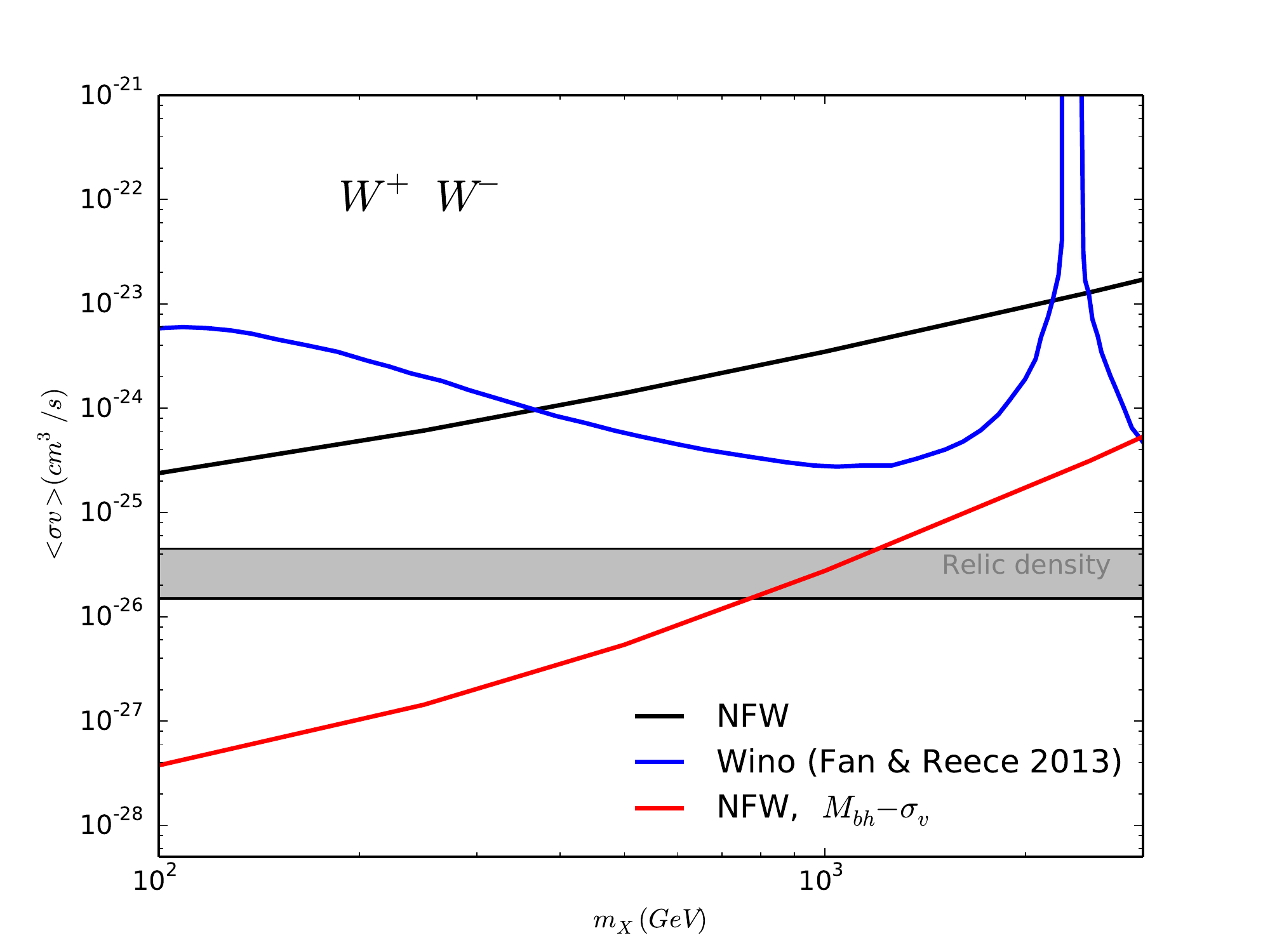}

 \vspace{-15pt}
  \caption{ DM annihilation cross-section constraints derived from a combined analysis of 15 dwarf spheroidal galaxies, for the $\tau \bar{\tau}$ (left) and $W\bar{W}$ final states. We used an NFW DM distribution, and a Tremaine relation to assign the black hole masses to each galaxy. }

\label{fig:Fermi3}
\end{figure*}

In figure \ref{fig:Fermi3} we show the combined constraints for two different annihilation pathways, namely the $\tau \bar{\tau}$ (left), and $W \bar{W}$ (right) final states. For these case we utilize a benchmark choice of a NFW density profile, and the Tremaine relation. As before, several dwarf galaxies, individually, would be excluding important regions of low mass WIMPs. For example, in the case of the $\tau \bar{\tau}$ channel it would be excluding again the Galactic center excess: we show the preferred region for the $\tau \bar{\tau}$ final state with a blue star, as calculated in Ref.~\cite{Abazajian:2014fta}. For the $W\bar{W}$ final state, the constraint we find would rule out wino dark matter, whose pair-annihilation cross section is shown with a solid blue line, as calculated in Ref.~\cite{Fan:2013faa}. 

\section{Discussion}

In this study we assessed how the presence of central black holes in local dwarf spheroidal galaxies impacts constraints on the dark matter pair-annihilation cross section from gamma-ray observations. Due to adiabatic contraction of the inner dark matter density distribution onto massive central objects, large central densities are generically predicted. The exceedingly steep profiles are cut off at some maximal density by annihilation processes, in a mass- and annihilation-rate-dependent fashion.

We explored three different prescriptions for the attribution of a mass for the black holes associated to a given dwarf galaxy. The various prescriptions stem from extrapolations of well-known relations between the velocity dispersion, or luminosity, of the stellar components of more massive galaxies with the central black hole mass. The three scenarios cover a wide range of possible black hole masses and, therefore, we expect our constraints to bracket a meaningful range of possible outcomes.

In contrast with previous analyses, and due to the maximal density cutoff alluded to above, the presence of a black hole enforces a correlation between the astrophysical  $J$ factor and the mass and pair-annihilation cross section of the dark matter particle. The most constraining dwarf galaxy depends upon the prescription for the attribution of the central black hole mass. Limits on the dark matter annihilation rate from observations of individual dwarfs are enhanced by many orders of magnitude in some cases.

We also attempted to derive a combined constraint that utilizes limits from observations of all 15 dwarfs in the original sample employed by the Fermi Collaboration. The joint constraint is always weaker than the constraint from the best single candidate, as a result of the wide spread in $J$ factors induced by the presence of a central black hole.

We find that taken at face value our results rule out a vanilla WIMP thermal relic for masses well in excess of 1 TeV for an NFW seed density profile, and of 100 GeV for a seed cored profile. A central black hole in local dwarfs would conclusively rule out dark matter annihilation as a source for the Galactic center gamma-ray excess, independently of the annihilation pathway, and would also solidly rule out wino dark matter.

We caution the Reader that there are several effects that could counteract the effect of the adiabatic contraction scenario we have presented. For instance, if the seed black hole was formed off-center \cite{Ullio:2001fb}, or if a stellar cusp around the black hole is present, inducing scattering of the dark matter particles  \cite{2004PhRvL..93f1302G},  the dark matter cusp could be smoothed out. Yet, the resulting density profile will be steeper than what considered in the Fermi analysis. The expected enhancement effect on the cross section constraints would be comparable to the more conservative scenarios we have discussed here.  
\acknowledgments

We thank Piero Madau, Joel Primack, Louis Strigari, and Octavio Valenzuela, for helpful comments and discussions. AXGM is supported by a UC MEXUS-CONACYT postdoctoral fellowship. FQ is supported in part by the Brazilian National Counsel for Technological and Scientic Development (CNPq). SP and FQ are supported in part by the US Department of Energy under Contract DE-SC0010107-001.

\begin{turnpage}
\begin{table}[h] \scriptsize
\caption{ \label{tab:dsphs} Properties of Milky Way dwarf spheroidal satellite galaxies.}

\begin{ruledtabular}
\begin{tabular}{ l c c c c c c c c}
Name                      & Distance &Luminosity& $\sigma_{*}$ & $\log_{10}({\rm J^{NFW}})$\footnotemark[1] &  $\log_{10}({\rm J^{Burkert}})$\footnotemark[1] & $m_{\rm bh}(\sigma_{*})$& $m_{\rm bh}(L_{*})$ &  \\
                          &  \rm kpc & $L_\odot$ &$\rm km/s$ & $\log_{10}[\rm GeV^2 \rm cm^{-5} \rm sr]$ & $\log_{10}[\rm GeV^2 \rm cm^{-5} \rm sr]$ & $M_{\odot}$ & $M_{\odot}$& \\
\hline
Bootes I                     & 66    & $3.0 \pm 0.6 *10^4$& 6.5 & $18.8 \pm 0.22$ & $18.6 \pm 0.17$  & $145.1$&$100.0$ \\
Canes Venatici I            & 218   & $7.9 \pm 3.6 *10^3$ & 7.6  & $17.7 \pm 0.26$ &  $17.6 \pm 0.17$ &$100.0$ & $100.0$\\
Carina                 & 105    & $2.4 \pm 1.0 *10^5$ & 6.6 & $18.1 \pm 0.23$ &  $18.1 \pm 0.16$ & $154.2$& $312.0$ \\
Coma Berenices            & 44 & $3.7\pm 1.7 *10^3$ &4.6 & $19.0 \pm 0.25$   & $18.9 \pm 0.21$  &100.0 &100.0\\
Draco                & 76     & $2.7 \pm 0.4 *10^5$&9.1 & $18.8 \pm 0.16$ &  $18.7 \pm 0.17$ &557.4 & 351.1\\
Fornax                 & 147    &$ 1.4\pm0.4 *10^7$ &11.7 &  $18.2 \pm 0.21$ & $18.1 \pm 0.22$ &1523.1  & $18200.0$\\
Hercules                 & 132   & $3.6\pm 1.1 *10^4$& 3.7& $18.1 \pm 0.25$   & $17.9 \pm 0.19$ &100.0&100.0 \\
Leo II                    & 233  &$5.9\pm 1.8 *10^5$ &6.6  & $17.6 \pm 0.18$ &  $17.5 \pm 0.15$ &154.2 & 767.0\\
Leo IV                  & 154    & $8.7\pm4.6 *10^3$&2.4 & $17.9 \pm 0.28$ &  $17.8 \pm 0.21$ &100.0   &100.0 \\
Sculptor                    & 86 &$1.4 \pm 0.6 *10^6$ &9.2    & $18.6 \pm 0.18$   & $18.5 \pm 0.17$ & 582.3  &1820.0\\
Segue 1                     & 23    & $3.3 \pm 2.1 *10^2$&4.3& $19.5 \pm 0.29$ & $19.4 \pm 0.24$ & 100.0  &100.0 \\
Sextans                     & 86 & $4.1 \pm 1.9 *10^5$&   7.9 & $18.4 \pm 0.27$   & $18.4 \pm 0.16$ &316.6  &533.0 \\
Ursa Major II             & 32     & $ 4.0\pm1.9 *10^3$&5.7 & $19.3 \pm 0.28$  & $19.2 \pm 0.21$ & 100.0 & 100.0 \\
Ursa Minor                 & 76     & $2.0 \pm 0.9 *10^5$& 9.5& $18.8 \pm 0.19$  & $18.7 \pm 0.20$ & 662.1 & 260.0\\
Willman 1            & 38     &$1.0 \pm 0.7 *10^3$ &4.3 & $19.1 \pm 0.31$ &   $19.0 \pm 0.28$ &100.0 &100.0  \\

\footnotetext[1]{ $J$ factors were calculated over a solid angle of $\Delta \Omega \sim 2.4 \times 10^{-4} \rm sr$.\cite{Ackermann:2013yva}}
\end{tabular}
\end{ruledtabular}

\end{table}
\end{turnpage}

\bibliographystyle{unsrtnat}
\bibliography{References}

\end{document}